\pdfoutput=1
\documentclass[11pt]{article}
\usepackage{graphicx,color} 
\usepackage{jheppub}
\usepackage{amsmath}
\usepackage{amssymb}
\usepackage{mathtools}
\usepackage[utf8]{inputenc}
\usepackage{parskip}
\DeclareMathOperator{\sech}{sech}
\newcommand{\beq}{\begin{equation}}
\newcommand{\eeq}{\end{equation}}

\def\o{\omega}

\def\f{\phi}
\def\rm{r_{-}}
\def\rp{r_{+}}

\def \O{\mathcal{O}}

\def\x{\mathbf{x}}
\def\k{\mathbf{k}}

\def \ti{\pi}

\def \betal_1{\frac{ (r_--r_+)-2  (r_-\tanh ^{-1}\frac{r}{{r_-}}-r_{+}\tanh ^{-1}\frac{r}{{r_{+}}})}{2(r^2_--r^2_+)}}
\newcommand{\nn}{\nonumber\\}

\title{Bulk reconstruction in 2D multi-horizon black hole}
\author[a]{Parijat Dey,}
\author[b]{Nirmalya Kajuri }
\author[b]{and Rhitaparna Pal}

\affiliation[a]{ Department of Astrophysics and High Energy Physics,\\
S.N. Bose National Centre for Basic Sciences,\\
Salt Lake, Kolkata 700106, India}
\affiliation[b]{School of Physical Sciences, Indian Institute of Technology, Mandi,\\ Kamand, Himachal Pradesh 175005, India}
\emailAdd{ parijat.dey@bose.res.in}
\emailAdd{nirmalya@iitmandi.ac.in}
\emailAdd{d21089@students.iitmandi.ac.in}

\abstract{The goal of the bulk reconstruction program is to construct boundary representations of fields in asymptotically Anti-de Sitter spacetimes. In this paper, we extend the program by computing the boundary representation of massless fields in an  Achucarro-Ortiz black hole spacetime. We obtain analytic expressions for smearing functions in both the exterior and interior of the black hole. We also obtain expressions for Papadodimas-Raju mirror operators.}

\begin{document}
\maketitle
\section{Introduction}
According to the AdS/CFT correspondence, all physical information of a $d+1-$dimensional Anti-de Sitter spacetime is fully encoded in a $d-$dimensional conformal field theory. Figuring out the complete dictionary between bulk and boundary is a key open problem.

A first step towards this goal is translating bulk fields in the bulk theory to operators in the boundary CFT. The program of mapping bulk fields to boundary operators has come to be known as bulk reconstruction\cite{Dobrev:1998md, Bena:1999jv,Hamilton:2005ju,Hamilton:2006az,Hamilton:2006fh,Heemskerk:2012np,Papadodimas:2012aq,Kabat:2011rz,Heemskerk:2012mn, Kabat:2012hp,Leichenauer:2013kaa,Sarkar:2014dma,Sarkar:2014jia,Guica:2014dfa,Roy:2015pga,Nakayama:2015mva,Kabat:2016rsx,Kabat:2017mun,Nakayama:2015mva,Kabat:2018pbj,Foit:2019nsr,Kajuri:2020bvi,Dey:2021vke,Bhattacharjee:2022ehq,Bhattacharjee:2023roq,Goldar:2024crc}. For a recent review, we refer to \cite{Kajuri:2020vxf}.

When considering the mapping between the entire bulk and the boundary, one finds a boundary representation of the form:
$$\phi(r,\x) = \int d^d \x' \,K(r,\x,\x') \O(\x') \,.$$

Here $(r,\x) $ are the  bulk coordinates and $\x$ are the boundary coordinates.

Here $K(r,\x,\x')$ is the smearing function, and the integration is over the entire boundary. This is known as HKLL construction in the literature\footnote{For subregion reconstructions, this applies for causal wedges. To reconstruct a bulk field outside of causal wedge but still in the entanglement wedge, a different construction is needed\cite{Faulkner:2017vdd,Cotler:2017erl}. We do not consider subregion reconstruction in this paper.}.

In this paper, we have constructed boundary representation for fields in a 2-dimensional black hole background. First, let us explain the motivation for this.

The question of bulk reconstruction in black hole backgrounds has been of interest for a long time. In AdS/CFT, the smoothness of the black hole horizon is equivalent to the existence of boundary representation for fields in the interior of a black hole\cite{Almheiri:2012rt,Heemskerk:2012mn}. For an eternal black hole, the boundary representations of fields inside the horizon can in principle be constructed via HKLL construction, with the resulting operator having support in both the boundary CFTs. For a collapsing black hole situation, where there is only one asymptotic boundary, HKLL construction does not work in the interior. Establishing horizon smoothness is more subtle in this case. One can still construct boundary representations for fields at late times by following the ``mirror operator'' construction of Papadodimas and Raju\cite{Papadodimas:2013jku,Papadodimas:2013wnh,Papadodimas:2015jra}. However, there is no  example of an analytic mirror operator smearing function in the literature.
 
Recently there has been much interest in the `island solution' of the black hole information paradox. While it is understood that information escapes the black holes, it is not entirely clear how. It would be helpful if one could track the information of fields in the bulk from the boundary theory at all times. In other to do this, one would need to know the boundary representations of bulk fields.(see for instance \cite{Chen:2019iro}). Explicit boundary representation formulae for black hole backgrounds would therefore be useful in developing a better understanding of information recovery from black holes.

Another interesting question pertains to the inner horizons of multi-horizon black holes. These inner horizons can be Cauchy horizons, in which case the evolution of fields beyond the inner horizon can not be determined from data at the inner horizon. It requires boundary conditions at the timelike singularities. But does the boundary CFT "know" how fields evolve beyond the inner horizon? It remains an open question if the region beyond the inner horizon is accessible to the boundary CFT(s)\cite{Balasubramanian:2019qwk}. Boundary representation of fields in multi-horizon spacetimes would be a helpful step in developing such an understanding.

Despite these points of interest and much work on the subject (see for instance \cite{Hamilton:2006fh,Papadodimas:2012aq,Leichenauer:2013kaa,Guica:2014dfa,Nakayama:2015mva,Kajuri:2020bvi}), analytic HKLL constructions of fields in the interior of a black hole are almost absent in the literature, even for double-sided black holes. Indeed, the only example is for pure AdS$_2$, viewed as a `black hole' in AdS-Rindler coordinates. The issue in dimensions greater than two is that the smearing function in a black hole background does not have an analytic expression but is distributional in nature \cite{Leichenauer:2013kaa,Morrison:2014jha}\footnote{One can get around this by complexifying the boundary\cite{Hamilton:2006fh}. In the complexified boundary approach, expressions for the smearing function in the interior have also been constructed. However, it is unclear if this is the right thing to do. There is no way to recover an expression on the real boundary in this approach (no analog of $i\epsilon$ prescription) and the physical meaning of working with a complex time, with regards to unitarity, is unclear.} This makes it harder to gain insights, although some progress can be made in reconstructing delocalised wave packets instead of fields\cite{Guica:2014dfa,Kajuri:2020bvi}. This problem, however, does not arise for 2-dimensional black holes and one can hope to find analytic representations of the smearing function in this case. 

All these points lead us to the problem considered in this paper: construction of boundary representation for Achucarro-Ortiz black holes. These are 2-dimensional asymptotically Anti-de Sitter black holes obtained by dimensional reduction from rotating BTZ black holes\cite{Achucarro:1993fd}(For an analysis on the stability of their Cauchy horizon, see \cite{Bhattacharjee:2020nul}). For this background, we obtained analytic expressions for boundary representations of minimally coupled massless fields\footnote{Field equations for massive fields in this background do not have analytic solutions. One can also consider the case of probe fields coupled with the dilaton, but this corresponds to the zero mode of the rotating BTZ black hole, and the smearing function can be read off from the rotating BTZ case\cite{Kajuri:2020bvi}.}. Analytic expressions for mirror operators are also obtained.

We have thus filled the gap in the literature for analytic expressions of boundary representations and mirror operators in a black hole background. The main results of the paper are given in \eqref{result1}, \eqref{result2} and \eqref{result3}. These should prove useful to answer the kind of questions we outlined above.

The paper is organized as follows. In  section \ref{bhback}, we provide a brief sketch of how boundary representations are constructed in black holes. Section \ref{brmassless} deals with the case of the massless scalar. Mirror operators for massless scalars are constructed in section \ref{mirrorc}. We conclude with a summary in \ref{conc}. 

\section{Bulk Reconstruction in Black Hole backgrounds}\label{bhback}

First, let us sketch the HKLL construction for pure AdS in $d$ spacetime dimensions. Consider a scalar field $\phi(r,\x)$ in the bulk. Let $\O(\x)$ be the CFT primary dual to $\phi(r,\x)$. From the extrapolate dictionary \cite{Banks:1998dd,Harlow:2011ke}, the correlation function of $\phi(r,\x)$ is related to that of $\O(\x)$ as:
\begin{align}
\notag \lim_{r\to \infty}r^{n\Delta} &\langle\phi(r,\x_1 )\phi(r,\x_2 )...\phi(r,\x_n )\rangle = \langle 0|\O(\x_1 )\O(\x_2 )..\O(\x_n )|0\rangle 
\end{align} 
where $\Delta$ is the conformal dimension of $\O$. Within correlation functions, we may then write:

\begin{align}\label{xp}
 \lim_{r\to \infty}r^{\Delta}  \phi(r,\x )= \O(\x) \,.
\end{align} 

The idea behind HKLL construction is to solve the bulk Klein-Gordon equation with \eqref{xp} as a ``boundary condition.'' 

One way of doing this in practice is to start from the mode expansion:
\beq \label{mode}
\phi(r,\x ) = \int d^dk\, \frac{1}{N_{\k}} f_{k}(r) e^{i \k \cdot \x}a_{\k } + \text{hermitian conjugate}\,,
\eeq
where $f_k(r)e^{i \k \cdot \x}$ are the \textit{normalizable} mode solutions of the Klein-Gordon equation and $N_\k$ is the normalization constant.
\begin{figure}
\centering
\includegraphics[width=0.8\linewidth]{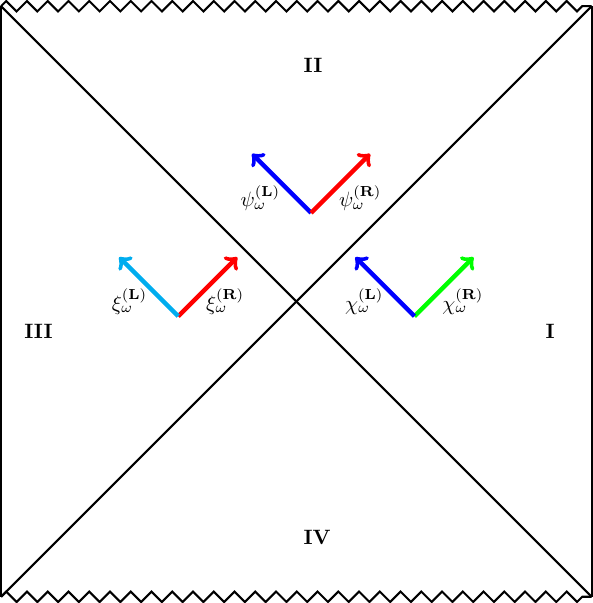}
\caption{Interior and exterior modes in a black hole. Left moving exterior modes $\chi^{(L)}_\o$ from region I continue into region II. Similarly right moving modes $\xi_\o^{(R)}$ from region III continue into region II. These match with the interior modes $\psi^{(L)}_\o,\psi^{(R)}_\o$ at the horizons.}
\label{bhfig}
\end{figure}
Substituting \eqref{xp} in \eqref{mode}, we first obtain boundary representations for the creation and annihilation operators: 

\beq \label{anni}
a_{\k}= \frac{N_{\k}}{M_{\k}}\int d^dx \, \O(\x)e^{-i\k\cdot\x}  \equiv \frac{N_{\k}}{M_{\k}}\O_\k\,,
\eeq
where 
\beq 
M_{\k}= \lim_{r\to \infty}r^{\Delta} f_{k }(r). 
\eeq

Then substituting \eqref{anni} in \eqref{mode}, one obtains the boundary representation of the bulk scalar, which can be expressed in the form:

\beq 
\phi(r,\x) = \int  d^dx^\prime\, K(r,\x;{\x^\prime})O({\x^\prime})
\eeq
where 
\beq 
K(r,\x ;{\x^\prime})=\int d^dk \,\frac{1}{M_{\k}} f_{k }(r) e^{i \k \cdot (\x -\x^\prime )}
\eeq 
is called the smearing function.

For an eternal black hole, there are two asymptotic boundary regions. In the exterior regions of the black hole (regions I and III in Figure \ref{bhfig}), the construction works exactly as in the case of pure AdS. So in region I, we will obtain:
\beq 
\phi_I(r,\x) = \int  d^dx^\prime\, K_I(r,\x;{\x^\prime})O({\x^\prime})\,,
\eeq
where 
\beq \label{smearI}
K_I(r,\x ;{\x^\prime})=\int d^dk\,\frac{1}{M_{\k}}f_{k }(r) e^{i \k \cdot (\x -\x^\prime )}\,,
\eeq 
and $f_k(r)$ are the normalizable modes that die down at the asymptotic boundary of I. 

The situation is different in the interior of the black hole (or white hole). The modes of the  region II $\psi_k(r,\x)=f_{k }(r) e^{i \k \cdot \x}$ can be written as a linear combination of left-moving and right-moving modes:
$$\psi_k(r,\x)= c_k\psi^{L}_k(r,\x)+d_k\psi^{R}_k(r,\x) \,.$$
The unknown constants are to be fixed by demanding that interior modes match the exterior normalizable modes at the horizon. This is illustrated in Figure \ref{bhfig}. As is evident from the figure, the left(right) moving modes from region I(III) cross the horizon and enter the interior region. 

Once $c_k,d_k$ are fixed, the boundary representation can be computed:
\begin{equation}
\phi_{II}(r,\x)= \int d^d\x^\prime K_L(r,x;x^\prime)\O_L(x^\prime) + \int d^d\x'' K_R(r,x;x'')\O_R(x'').
\end{equation}

where
\begin{equation}\label{smear2}
 K_R(r,x;x^\prime)= \int d^d\k\, \frac{c_k}{M_k} \psi^{R}_k(r,\x)e^{-i\k\cdot\x^\prime}+\text{h.c}. 
\end{equation}
Here $\O_L,\O_R$ refer to operators on the left and right CFT respectively. 

$K_L(r,x;x^\prime)$ is obtained similarly, or directly from $K_R$ via a CPT transformation. The boundary representation for $\phi_{II}$ is a sum of operators on left and right CFT. 

One can also find a boundary representation with support on only one boundary, using the Papadodimas-Raju mirror operator construction. 

For a CFT state $|\psi\rangle$ dual to a two-sided black hole, the starting point of the mirror operator construction is the observation\cite{Papadodimas:2015jra}:

\begin{align}
\O^L_\k |\psi\rangle &= e^{-  \omega}\left(\O^R_\k\right)^\dagger |\psi\rangle\,,\\
\left(\O^L_\k\right)^\dagger |\psi\rangle &= e^{  \omega}\O^R_\k |\psi\rangle\,,
\end{align}
where $\omega$ is the time component of $\k$. 

This allows us to translate between $\O_L$ and $\O_R$ as long as we are working with the state $|\psi\rangle$(or small excitations around it). We can then find a boundary representation supported in only boundary (say, right):
\beq 
\phi_{II} =\int d^d\x^\prime \, \left(K_R(r,x;x^\prime)\O_R(x^\prime)+ K_\text{mirror}(r,x;x^\prime)\O_R(x^\prime) \right).
\eeq
where 
\beq \label{mirror}
K_\text{mirror}(r,x;x^\prime)=\int d^dk\,\frac{d_k}{M_k} \left(e^{  \omega}\psi^{L}(r,\x)e^{-i\k \cdot\x}+ e^{-  \omega}\left(\psi^{L}(r,\x)\right)^*e^{i\k \cdot\x}\right)\,.
\eeq

The mirror operator construction is state-dependent. This representation will be valid on a subspace of the CFT that contains the state $|\psi\rangle$ and small excitations of $|\psi\rangle$. This is the subspace corresponding to bulk effective field theory on this background.

For collapsing black holes, regions III and IV are absent. In this case, HKLL construction fails for the interior of the black hole\cite{Harlow:2014yka}. However, the mirror operator construction provides a boundary representation that is valid at late times. 

\section{Boundary Representation for Massless Scalar }\label{brmassless}

Now we will use the formulae from the previous section to obtain boundary representations for the Achucarro-Ortiz black hole.

\begin{figure}
\centering
\includegraphics[width=0.5\linewidth]{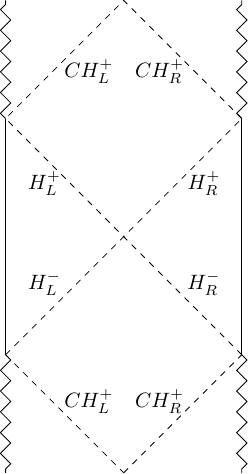}
\caption{Penrose diagram of Achucarro-Ortiz black hole. $H_L^+,H_R^+$ are the future right and left horizons respectively. $H_L^-,H_R^-$ are the past right and left horizons. $CH_L^+,H_R^+$ are the left and right inner horizons.}
\label{aofig}
\end{figure}

The metric of an Achucarro-Ortiz black hole is given by\cite{Achucarro:1993fd}:
\begin{align}\label{met}
    ds^2 &=-F(r)dt^2+\frac{dr^2}{F(r)}\,,\nn
    F(r)&=\frac{(r^2-r^2_+)(r^2-r^2_-)}{r^2}
\end{align}
Here $r_+,r_-$ denote the outer and inner horizons respectively. The Penrose diagram is shown in Figure \ref{aofig}.

The massless Klein-Gordon equation
\begin{align}
\Box \Phi=0\,,
\end{align}

translates, after substituting $\Phi(r,t)=e^{i \omega t} \f_{\omega}(r)$, to the following equation: 
\beq 
F(r) \frac{\partial }{\partial r}\left(F(r) \frac{\partial f_\omega(r)}{\partial r}\right)+\omega^2 f_\omega(r)=0
\eeq

This equation has the general solution:
\begin{align} \notag
f_\omega(r)=A_\o \cos &\left(\frac{\o \left({\rm} \tanh ^{-1}\left(\frac{r}{{\rm}}\right)-{\rp} \tanh ^{-1}\left(\frac{r}{{\rp}}\right)\right)}{({\rm}-{\rp}) ({\rm}+{\rp})}\right)\\ \label{fullsol} &+B_\o \sin \left(\frac{\o \left({\rm} \tanh ^{-1}\left(\frac{r}{{\rm}}\right)-{\rp} \tanh ^{-1}\left(\frac{r}{{\rp}}\right)\right)}{({\rm}-{\rp}) ({\rm}+{\rp})}\right)
\end{align}
where $A_\o,B_\o$ are undetermined constants.
\subsection{Region I}
To obtain the boundary representation for a scalar in region I, we only need the normalizable mode solution. Demanding that the solution vanishes as $r\to \infty$ fixes the solution (up to normalization constant) to be

\begin{align}\label{reg1sol}
    f^{norm}_\o(r)& = \sech \frac{  \omega }{2(r_++r_-)}  \sin \left(\o\frac{ (r_--r_+)-2  (r_-\tanh ^{-1}\frac{r}{{r_-}}-r_{+}\tanh ^{-1}\frac{r}{{r_{+}}})}{2(r^2_--r^2_+)}\right)
    \end{align}
  For a massless field in $d=2 $, $\Delta=1$.
Taking the boundary limit, of the above, we find:
\beq 
M_\o= \lim_{r \to \infty} r f_\o(r)=\o\sech  \frac{  \omega }{2(r_++r_-)} 
\eeq
 We can now use \eqref{smearI} to find the smearing function for this region. This gives:

\begin{align}
  K_I(r;t,t')=  \int_{-\infty}^{\infty} d\omega \, \frac{1}{{\omega}} {\sin \left(\omega  \betal_1\right)} \cos{ \omega (t-t')}\,.
\end{align}

Performing this integral, one obtains the expression: 
\beq \label{result1}
  K_I(r;t,t')=\frac{\ti}{4}\theta \left(\betal_1-(t-t^\prime)\right)\,.
\eeq 

The expression for the boundary representation of a massless bulk field in region I is: 
\beq \label{efsmear}
\phi(r,t)= \frac{\ti}{4}\int dt^\prime\,\theta \left(\betal_1-(t-t^\prime)\right)\O(t^\prime)
\eeq

If we rewrite the smearing function in terms of the Eddington-Finkelstein radial coordinate $r_*$, which is defined via

\begin{equation} \label{efr}
dr_* = \frac{dr}{F(r)} \implies r_* = \frac{{\rm} \tanh^{-1}\left( \frac{r}{{\rm}} \right) - {\rp} \tanh^{-1}\left( \frac{r}{{\rp}} \right)}{({\rm} - {\rp}) ({\rm} + {\rp})}
\end{equation}

then the smearing function can be written as:
\beq 
K_I(r;t,t')=\frac{\ti}{4}\theta\left(\frac{  \o}{2(\rp+\rm)}-(r_*+(t-t^\prime)) \right)\,.
\eeq

This form provides some insight into the smearing function. For pure AdS, the smearing function is found to have support only at spacelike separated points on the boundary, that is at $r_*>t$. However, in this case, we find that the support is shifted by a constant factor of $\frac{  \o}{2(\rp+\rm)}$. Such shift in the support was also found in the case of the (approximate)smearing function for rotating BTZ black hole\cite{Kajuri:2020bvi}.

\subsection{Region II}
To construct the boundary representation in region II, it is useful to use \eqref{fullsol} and write

 \begin{align}\notag
f_\omega(r) = C_\omega &e^{i \left( \frac{\omega \left( {\rm} \tanh^{-1} \left( \frac{r}{{\rm}} \right) - {\rp} \tanh^{-1} \left( \frac{r}{{\rp}} \right) \right)}{({\rm} - {\rp}) ({\rm} + {\rp})} \right)} \\
&+ D_\omega e^{-i \left( \frac{\omega \left( {\rm} \tanh^{-1} \left( \frac{r}{{\rm}} \right) - {\rp} \tanh^{-1} \left( \frac{r}{{\rp}} \right) \right)}{({\rm} - {\rp}) ({\rm} + {\rp})} \right)}
\label{fullsol2}
\end{align}

These correspond to the left and right moving modes once the $e^{i \o t}$ factor is multiplied. 

The constants $C_\o,D_\o$ are determined by matching the solution at the horizon to the normalizable modes in the exterior. For the horizon between region I and II, the interior modes must match with the normalizable modes in I at the horizon. 

The normalizable mode in region I given in \eqref{reg1sol} can also be written as a combination of the left and right moving modes:

The constants $C_\o,D_\o$ are determined by matching the solution at the horizon to the normalizable modes in the exterior. For the horizon between region I and II, the interior modes must match with the normalizable modes in I at the horizon. 

The normalizable mode in region I given in \eqref{reg1sol} can also be written as a combination of the left and right moving modes: 
\begin{align}\notag
f^{norm}_\o(r)= \tan \frac{  \o}{2(\rp+\rm)}&\left(\left(\frac{1}{2}+\frac{1}{2i}\right)e^{i\frac{\o \left({\rm} \tanh ^{-1}\left(\frac{r}{{\rm}}\right)-{\rp} \tanh ^{-1}\left(\frac{r}{{\rp}}\right) \right)}{({\rm}-{\rp})}}\right. \\ \label{fi} &+\left.\left(\frac{1}{2}-\frac{1}{2i}\right)e^{-i\frac{\o \left({\rm} \tanh ^{-1}\left(\frac{r}{{\rm}}\right)-{\rp} \tanh ^{-1}\left(\frac{r}{{\rp}}\right) \right)}{({\rm}-{\rp}}}\right)\,.
\end{align}

The left moving modes match at the horizon\footnote{The way to see this technically is to write the modes as a function Kruskal coordinates $U,V$. The right moving modes $U^{i\o}$ drop out of the mode expansion at the horizon $U \to 0$ between I and II and one needs to match only the left moving modes. See the appendix A2 of \cite{Papadodimas:2012aq} for a detailed example.}
Comparing \eqref{fullsol} and \eqref{fi}, we fix the constant:
\beq 
C_\o=\tan \frac{  \o}{2(\rp+\rm)}\left(\frac{1}{2}+\frac{1}{2i}\right)\,.
\eeq

We now use compute the smearing function $K_R$ supported in the asymptotic boundary of region I using \eqref{smear2}:

\beq 
K_R(r,t,t^\prime) = \int d\o\, \frac{1}{\o}\sin \frac{  \o}{2(\rp+\rm)}\left(\left(\frac{1}{2}+\frac{1}{2i}\right)e^{i\left(\frac{\o \left({\rm} \tanh ^{-1}\left(\frac{r}{{\rm}}\right)-{\rp} \tanh ^{-1}\left(\frac{r}{{\rp}}\right) \right)}{({\rm}-{\rp})} +t-t^\prime\right)}+h.c \right)\,.
\eeq
This simplifies to:
\begin{align}\notag
K_R(r,t,t^\prime) = \int d\o\, \frac{1}{\o}&\sin \frac{  \o}{2(\rp+\rm)}\left(\cos{\left(\frac{\o \left({\rm} \tanh ^{-1}\left(\frac{r}{{\rm}}\right)-{\rp} \tanh ^{-1}\left(\frac{r}{{\rp}}\right) \right)}{({\rm}-{\rp})} +t-t^\prime\right)} \right.\\ &\left.-\sin {\left(\frac{\o \left({\rm} \tanh ^{-1}\left(\frac{r}{{\rm}}\right)-{\rp} \tanh ^{-1}\left(\frac{r}{{\rp}}\right) \right)}{({\rm}-{\rp})} +t-t^\prime\right)} \right)\,.
\end{align}

Performing the above integration, we finally get the result: 
\begin{align}\label{result2}
  K_R(r,t,t^\prime) =\frac{\ti}{4}\theta\left(\frac{  \o}{2(\rp+\rm)}-(r_*+(t-t^\prime))\right)+\frac{1}{2}\log \left(\frac{\frac{  \o}{2(\rp+\rm)}+(r_*+(t-t^\prime))}{\frac{  \o}{2(\rp+\rm)}-(r_*+(t-t^\prime))}\right)\,,
\end{align}
where $r_*$ is given by \eqref{efr}. $K_L(r,t,t^\prime)$ can be obtained by transforming $t\to -t, t^\prime \to -t^\prime$.

We find that the first term in the interior smearing function has the same form as the exterior smearing function. However, a new logarithmic term appears in this case. 

%%%%%%%%%%%%%%%%%%%%%%%%%%%%%%

 \section{Mirror Operator Construction}\label{mirrorc}

 If we consider a scenario where an Achucarro-Ortiz black hole is formed via collapse, then only one asymptotic boundary will be present. In this case, the HKLL construction above fails. However, the Papadodimas-Raju mirror operator construction can be used to obtain a state-dependent boundary representation which will be valid at late times. 

Let us consider the case where we have only regions I and II. To find the mirror we first fix the constant $C_\o$ in \eqref{fullsol2} in the same way as above. 

Then using the formula \eqref{mirror}, we find the smearing function:
\begin{align}\notag
K_\text{mirror}(r,x;x^\prime)=\int d\o\,\frac{1}{\o}\sin \frac{  \o}{2(\rp+\rm)}& \left(e^{\ti \omega} \left(\frac{1}{2}+\frac{1}{2i}\right)e^{i\o(r_*-(t-t^\prime))} \right.\\& \left. +e^{-\ti \omega}\left(\frac{1}{2}-\frac{1}{2i}\right)e^{-i\o(r_*-(t-t^\prime))}\right)
 \end{align}
This also turns out to have an analytic expression:
\begin{align}\notag
K_\text{mirror}(r,x;x^\prime)=\frac{\ti}{4}\theta\left(\frac{  \o}{2(\rp+\rm)}-(r_*-(t-t^\prime))\right)+\frac{1}{2}\log \left[\frac{\frac{  \o}{2(\rp+\rm)}+(r_*-(t-t^\prime))}{\frac{  \o}{2(\rp+\rm)}-(r_*-(t-t^\prime))}\right]\\ \notag +\log\left[1+\left(\frac{\ti}{\frac{  \o}{2(\rp+\rm)}+(r_*-(t-t^\prime))}\right)^2 \right]-\log\left[1-\left(\frac{\ti}{\frac{  \o}{2(\rp+\rm)}-(r_*-(t-t^\prime))}\right)^2\right]\\-2 \tanh^{-1}\left(\frac{\ti}{\frac{  \o}{2(\rp+\rm)}+(r_*-(t-t^\prime))}\right)-2 \tanh^{-1}\left(\frac{\ti}{\frac{  \o}{2(\rp+\rm)}-(r_*-(t-t^\prime))}\right)
\label{result3}
\end{align}

The first two terms have the same form as \eqref{result2} but there are several other terms. 

Thus, for an Achucarro-Ortiz black hole formed via collapse, the boundary representation at late times will be given by:
\beq 
\phi^{II}(r,t)= \int dt' \left(K_R(r,t;t^\prime) \O_R(t^\prime) + K_{\text{mirror}}(r,t,t^{\prime})\O_R(t^\prime)\right)
\eeq
with $K_R(r,t;t^\prime)$ given by \eqref{result2}and $K_{\text{mirror}}(r,t,t^{\prime})$ given by \eqref{result3}.

\section {Summary}\label{conc}

In this paper, we constructed boundary representations for a massless scalar in an Achucarro-Ortiz black hole background. As noted in the introduction, there are several open questions in black hole physics for which boundary representations of bulk fields should be helpful. However, analytic expressions for smearing functions in black hole backgrounds are scarce. Our result fills a gap in the literature by providing such an example. 

We considered a massless free scalar and obtained its boundary representation following the HKLL construction. By employing the mode sum approach, we constructed the smearing functions in both the interior and exterior of the black hole. The results are given in \eqref{result1} and \eqref{result2} respectively. We find that while the smearing function in the exterior has the familiar form of a $\theta$ function, an additional logarithmic term appears appears in the smearing function for the interior region. 

In the case of a collapsing black hole, there is only one asymptotic boundary and the HKLL construction fails for the interior. However, one can still construct a boundary representation by following the Papadodimas-Raju mirror operator construction. We followed this construction and found an analytic expression for the corresponding smearing function, given in \eqref{result3}. Thus we obtained boundary representations for a massless scalar for both eternal and collapsing Achucarro-Ortiz black holes.

It is straightforward to generalize the results to free gravitons. In the holographic gauge, a graviton field can be recast as a set of free massless scalars\cite{Kabat:2012hp} and our results will apply. 

In the future, we will be interested in applying our results to answer some of the outstanding questions about black hole physics in holography.

\begin{acknowledgments}
NK is supported by SERB Start-up Research Grant SRG/2022/000970.
\end{acknowledgments}

%%%%%%%%%%%%%%%%%%%%%%%%%%%%%%%%%%%%%

\bibliographystyle{JHEP}
\bibliography{adsrindler}

\end{document}